\documentclass[11pt,preprint]{aastex}
\pdfoutput=1
\usepackage{bm}
\usepackage{graphicx}
\usepackage{footnote}
\usepackage{color}
\usepackage{amsmath}
\usepackage{mathtools}
\usepackage{pdflscape}
\usepackage{multirow}

\newcommand{\mubold}{\bm{\mu}}
\newcommand\ltsima{$\; \buildrel <\over\sim \;$}
\newcommand\simlt{\lower.5ex\hbox{\ltsima}}
\newcommand\gtsima{$\; \buildrel >\over\sim \;$}
\newcommand\simgt{\lower.5ex\hbox{\gtsima}}

\newcommand\msun {M_\odot}
\newcommand\mearth {{M_\oplus}}

\shorttitle{}
\shortauthors{Bhattacharya et al}

\begin{document}

\title{WFIRST Exoplanet Mass Measurement Method Finds a Planetary Mass of
$39\pm 8 M_\oplus$ for OGLE-2012-BLG-0950Lb}


\author{A.~Bhattacharya\altaffilmark{1,2},
 J.P.~Beaulieu \altaffilmark{3, 4}, D.P.~Bennett\altaffilmark{1,2}, J.~Anderson\altaffilmark{5}, N.~Koshimoto\altaffilmark{6,7}, J.~R.~Lu\altaffilmark{8}\\ and \\ V.~Batista\altaffilmark{3}, J.~W.~Blackman\altaffilmark{4}, I.~A.~Bond\altaffilmark{9}, A.~Fukui\altaffilmark{10}, C.~B.~Henderson\altaffilmark{11}, Y.~Hirao\altaffilmark{2,12} J.~B.~Marquette\altaffilmark{3}, P.~Mroz\altaffilmark{13}, C.~Ranc\altaffilmark{14}, A.~Udalski\altaffilmark{13}}
\keywords{gravitational lensing: micro, planetary systems}

\altaffiltext{1}{Code 667, NASA Goddard Space Flight Center, Greenbelt, MD 20771, USA\\
  Email: {\tt aparna.bhattacharya@nasa.gov}}
\altaffiltext{2}{Department of Astronomy,
    University of Maryland, College Park, MD 20742, USA}   
\altaffiltext{3}{UPMC-CNRS, UMR 7095, Institut d’Astrophysique de Paris, 98Bis Boulevard Arago, F-75014 Paris}
\altaffiltext{4}{School of Physical Sciences, University of Tasmania, Private Bag 37 Hobart, Tasmania 7001 Australia}
\altaffiltext{5}{Space Telescope Institute, 3700 San Martin Drive, Baltimore, MD 21218, USA}
\altaffiltext{6}{Department of Astronomy, Graduate School of Science, The University of Tokyo, 7-3-1 Hongo, Bunkyo-ku, Tokyo 113-0033, Japan}
\altaffiltext{7}{National Astronomical Observatory of Japan, 2-21-1 Osawa, Mitaka, Tokyo 181-8588, Japan}
\altaffiltext{8}{University of California Berkeley, Berkeley, CA}
\altaffiltext{9}{Institute of Natural and Mathematical Sciences, Massey University, Auckland 0745, New Zealand}
\altaffiltext{10}{Okayama Astrophysical Observatory, National Astronomical Observatory of Japan, Asakuchi,719-0232 Okayama, Japan}
\altaffiltext{11}{NASA Exoplanet Science Institute, IPAC/Caltech, Pasadena, California 91125, USA}
\altaffiltext{12}{Department of Earth and Space Science, Graduate School of Science, Osaka University, 1-1 Machikaneyama, Toyonaka, Osaka 560-0043, Japan}
\altaffiltext{13}{Warsaw University Observatory, Al.~Ujazdowskie~4, 00-478~Warszawa, Poland}
\altaffiltext{14}{Code 660, NASA Goddard Space Flight Center, Greenbelt, MD 20771, USA}

\begin{abstract}
We present the analysis of the simultaneous high resolution images from the {\it Hubble Space Telescope} 
and Keck Adaptive Optics system of the planetary event OGLE-2012-BLG-0950 that determine that the system
consists of a $0.58 \pm 0.04 \msun$ host star orbited by a $39\pm 8\mearth$ planet of at projected separation of
$2.54 \pm 0.23\,$AU. The planetary system is located at a distance of $2.19\pm 0.23$ kpc from Earth.  This
is the second microlens planet beyond the snow line with a mass measured to be in the mass range 
$20$--$80 \mearth$. The runaway gas accretion process of the core accretion model predicts few planets 
in this mass range, because giant planets are thought to be growing rapidly at these masses and they 
rarely complete growth at this mass. So, this result suggests that the core accretion theory may need revision. 
This analysis also demonstrates the techniques that will be used to measure the masses of planets and 
their host stars by the WFIRST exoplanet microlensing survey: one-dimensional microlensing parallax 
combined with the separation and brightness measurement of the unresolved source and host stars to 
yield multiple redundant constraints on the masses and distance of the planetary system.  
\end{abstract}

\section{Introduction}
\label{sec-intro}
Gravitational microlensing is currently the only technique to detect planets just outside the snow line 
\citep{gouldloeb1992}  with masses as low as that of the Earth \citep{bennett96}. This method has so far 
discovered about 70 planets. For most events, the light curve modeling of these microlensing exoplanets 
provides the planet-star mass ratio, but it does not provide the masses for either the planet or the host
star. \citet{bennett07} showed theoretically that we can detect the foreground lens (host star and the planet) and 
the background source separately with high angular resolution follow-up observations taken a few years after the peak
magnification, and these observations can determine the host star mass and distance. The planet mass
is then determined from the microlensing light curve determination of the planet-star mass ratio.
High angular resolution follow-up observations of the planetary microlensing event OGLE-2005-BLG-169
using the {\it Hubble Space Telescope} (HST) \citep{ogle169} and Keck adaptive optics (AO) 
system \citep{batistaogle169} demonstrated this method, and measured the masses and distance
of this planetary system. These observations also confirmed the planetary interpretation of the
microlensing light curve because the lens-source relative proper motion predicted from the planetary signal was consistent with the one measured by the follow-up observations. 

In addition to this event, there are a number of planetary microlensing events that have had 
excess starlight detected at the position of the source in the high angular resolution follow-up 
observations, however very few managed to measure the lens-source separation. Under the assumption 
that this excess flux is due to the planetary host star, a number of papers have claimed to 
determine the host star mass \citep{janczak10,moa192_naco,batista2014}, but further follow-up
observations by \citet{moa310} have indicated that the excess flux for one of these events
was not due to the host star. A detailed analysis by \citet{kosh17_mb16227} showed that
excess star light that is unresolved from the source can often be due to stars other than the 
lens, such as companions to the source or lens, or unrelated stars. In cases where we were
able to measure the microlensing parallax, we do not require confirmation of the predicted lens-source
relative motion. Instead, high angular resolution follow-up observations
that don't resolve the lens and source stars can confirm and refine the microlensing parallax
mass measurement \citep{gaudi-ogle109,bennett2010,ogle0026} or distinguish between degenerate light curve
models \citep{bennett16,sumi16}.

This paper presents the first result from the NASA Keck Key Strategic Mission Support (KSMS) program, in
support of WFIRST, entitled ``Development of the WFIRST Exoplanet Mass Measurement Method\rlap,"
with follow-up observations of planetary microlensing event OGLE-2012-BLG-0950. This analysis 
demonstrates all of the methods that are expected to be the major host star and planet measurement
methods for the WFIRST exoplanet microlensing survey \citep{bennett07}. The HST observations are
part of a pilot program to use near simultaneous observations with Keck and HST to measure the separation 
of the source and planet host stars using the color dependent centroid shift method \citep{bennett06}
in the optical and infrared. The Keck and HST images can be used separately to measure the source-host
star separation using the image elongation method \citep{ogle169}. With the image elongation method, the 
lens and the source are partially resolved such that their total point spread function (PSF) is substantially 
elongated. So, in order to detect the lens we need to fit multiple stellar profiles to the target. 
However, if the PSF is not substantially elongated, we can still detect the lens from the color-dependent 
centroid shift method. In this case, when the lens and the source have very different colors, their relative 
brightness is very different in different passbands. As a result, the combination of the source+lens 
flux will have different centroids in different passbands at the same time. Hence, observing the target nearly 
simultaneously in three different passbands will give three different centroids for the same target. 
The shift of these centroids between different passbands can yield the separation and hence the 
detection of the lens. To demonstrate this method, we took a near simultaneous observation of 
the event OGLE-2012-BLG-0950 with HST $I$, $V$ and Keck $K$ passbands. 

The microlensing event OGLE-2012-BLG-0950 \citep{koshimoto950} was observed by the microlensing 
survey telescopes of the Optical Gravitational Lensing Experiment (OGLE) and the Microlensing Observations
in Astrophysics (MOA) collaborations. The anomaly at HJD$'=$6149 was observed primarily in the MOA data,
in addition to a single OGLE observation during the anomaly. No significant finite source effect from the 
background source was detected 
in the light curve modeling, but a significant microlensing parallax signal was seen. However, as is often the
case \citep{muraki11}, only a single component of the parallax vector ($\pi_{\rm E,E}$) was well measured
with any precision. The second component ($\pi_{\rm E,N}$) is constrained only with an upper limit
on $|\pi_{\rm E,N}|$. \citet{koshimoto950} attempted to determine the mass of the lens system with the
microlensing parallax measurement and the excess flux seen at the position of the source in 2013 AO follow-up
imaging with the Keck NIRC2 camera, but the uncertainty in the $\pi_{\rm E,N}$ and the possibility that
some or all of the excess flux could be due to a star other than the planetary host star, the masses of 
the host star and its planet remained uncertain.
 
In this paper, we present the first planetary microlensing event in which the lens-source separation is 
measured, which allows us to convert the one-dimensional microlensing parallax measurement to 
a complete microlensing parallax measurement \citep{ghosh04}. We combine this with the lens-source separation,
measured in three passbands,
to obtain a direct measurement of the lens system mass \citep{gould-lmc5}.
The paper is organized as follows: Section \ref{sec-lc_model} discusses the light curve modeling done to extract the light
curve constraints to be combined with the followup observations to determine the properties of the
lens system.Section \ref{sec-Followup} describes the details of our high resolution 
follow up observations and their photometry calibrations. Section \ref{sec-hst} focuses on the HST 
astrometric and photometric analyses with single star and dual star PSF fits of the blended source plus 
lens target. Section \ref{sec-Keck} explores the astrometric and photometric analyses of Keck AO 
images. Section \ref{sec-centroid} describes the color dependent centroid shifts between the positions 
of the blended source plus lens target in different passbands. In sections \ref{sec-murel} and \ref{sec-parallax}, 
we determine the geocentric relative lens-source proper motions and show that the identification 
of the lens constrains the parallax vector. Finally in sections \ref{sec-lens} and \ref{sec-discussion}, 
we discuss the lens exoplanet system properties and its implications.   

\begin{deluxetable}{cccccc}
\tablecaption{Best Fit Unconstrained Model Parameters
                         \label{tab-UCmparams} }
\tablewidth{0pt}
\tablehead{
& \multicolumn{2}{c} {$u_0 < 0$} & \multicolumn{2}{c} {$u_0 > 0$} &  \\
\colhead{parameter}  & \colhead{$s<1$} & \colhead{$s\approx 1$} & \colhead{$s<1$} & \colhead{$s\approx 1$} &\colhead{MCMC averages}
}  
\startdata
$t_E$ (days) & 70.823 & 71.059 & 70.093 & 70.062 & $69.8\pm 2.0$  \\   
$t_0$ (${\rm HJD}^\prime$) & 6151.4951 & 6151.4988 & 6151.4805 & 6151.4785 & $6151.492\pm 0.028$  \\
$u_0$ & -0.09764 & -0.09734 & 0.09634 & 0.09648 & $0.004\pm 0.098$  \\
$s$ & 0.89915 & 1.00082 & 0.89937 & 1.00243 & $0.942 \pm 0.056$  \\
$\alpha$ (rad) & -1.94169 & -1.94301 & 1.94842 & 1.94873 & $-0.07\pm 1.95$  \\
$q \times 10^{4}$ & 1.6469 & 1.5668 & 1.6669 & 1.6295 & $1.95 \pm 0.38$  \\
$t_\ast$ (days) & 0.03932 & 0.01433 & 0.01943 & 0.01324 & $0.057\pm 0.030$ \\
$\pi_{\rm E,N}$ & -0.2364 & -0.2470 & -0.2040 & -0.2020 & $-0.09\pm 0.22$ \\
$\pi_{\rm E,E}$ & -0.1190 & -0.1179 & -0.0988 & -0.1023 & $-0.122\pm 0.028$\\
$I_s$ & 19.299 & 19.303 & 19.315 & 19.314 & $19.253\pm 0.038$  \\
$V_s$ & 20.807 & 20.810 & 20.823 & 20.821 & $20.762\pm 0.038$  \\
fit $\chi^2$ & 6488.84 & 6490.48 & 6488.99 & 6490.60 &  \\
\enddata
\end{deluxetable}

\begin{deluxetable}{cccccc}
\tablecaption{Best Fit Model Parameters with $\mubold_{\rm rel}$ and Magnitude Constraints
                         \label{tab-Cmparams} }
\tablewidth{0pt}
\tablehead{
& \multicolumn{2}{c} {$u_0 < 0$} & \multicolumn{2}{c} {$u_0 > 0$} &  \\
\colhead{parameter}  & \colhead{$s<1$} & \colhead{$s\approx 1$} & \colhead{$s<1$} & \colhead{$s\approx 1$} &\colhead{MCMC averages}
}  
\startdata
$t_E$ (days) & 68.007 & 67.628 & 68.919 & 68.995 & $68.1\pm 1.2$  \\   
$t_0$ (${\rm HJD}^\prime$) & 6151.4702 & 6151.4749 & 6151.4978 & 6151.4999 & $6151.484\pm 0.027$  \\
$u_0$ & -0.09968 & -0.09734 & 0.10088 & 0.10073 & $0.043\pm 0.091$  \\
$s$ & 0.89783 & 1.00136 & 0.89791 & 0.99900 & $0.928 \pm 0.052$  \\
$\alpha$ (rad) & -1.94592 & -1.94719 & 1.93987 & 1.94146 & $-0.84\pm 1.75$  \\
$q \times 10^{4}$ & 1.7255 & 1.7442 & 1.6726 & 1.6165 & $2.01 \pm 0.39$  \\
$t_\ast$ (days) & 0.03634 & 0.03633 & 0.03609 & 0.01324 & $0.0366\pm 0.0013$ \\
$\pi_{\rm E,N}$ & 0.2107 & 0.2103 & 0.2192 & 0.2178 & $0.213\pm 0.017$ \\
$\pi_{\rm E,E}$ & -0.1536 & -0.1536 & -0.1685 & -0.1672 & $-0.157\pm 0.016$\\
$I_s$ & 19.274 & 19.266 & 19.260 & 19.262 & $19.265\pm 0.023$  \\
$V_s$ & 20.783 & 20.773 & 20.768 & 20.769 & $20.734\pm 0.023$  \\
fit $\chi^2$ & 6490.25 & 6491.87 & 6491.88 & 6493.45 &  \\
\enddata
\end{deluxetable}

\section{Light Curve Models}
\label{sec-lc_model}

Our analysis includes constraints on the lens system from both the microlensing light curve and
the high angular resolution follow-up observations, and we have found it most convenient to
redo the light curve analysis of this event that was previously presented by \citet{koshimoto950}.
Our re-analysis uses the same data set used by  \citet{koshimoto950} except that the MOA-II
survey light curve data have been re-reduced using the procedure described in \citet{bond17}.
This re-reduction procedure includes a light curve detrending procedure that is designed to remove
systematic photometry errors due to the differential refraction of neighboring stars \citep{bennett12},
as well as other seeing and air mass effects.

The modeling was done with the image centered ray shooting method \citep{bennett96,bennett-himag},
and the results are summarized in Table~\ref{tab-UCmparams}. The parameters of these models can be
separated into several categories. There are three parameters that are required for single lens light 
curves: the Einstein radius crossing time, $t_E$, the time of closest lens-source alignment, $t_0$, and
the lens source separation at the time of closest alignment, $u_0$, which is dimensionless because it
is given in units of the Einstein radius. For a binary lens system, $t_0$ and $u_0$ refer to the time of 
closest alignment between the source and the lens system center-of-mass. Four additional parameter
are generally included to describe binary lens systems. These are: the mass ratio, $q$, between the two
lens masses, the angle, $\alpha$, between the source trajectory and the lens axis, the separation,
$s$, between the two lens masses, in units of the Einstein radius, and the source radius crossing time,
$t_*$, which is needed for most planetary events because the sharp planetary light curve features often
resolve the finite angular size of the source star. Finally, there are the two components of the 
microlensing parallax vector, $\pi_{\rm E,N}$  and $\pi_{\rm E,E}$, which describe the effect of the
orbital motion of the observers on the Earth around the Sun.

As explained in the discovery paper,
\citep{koshimoto950}, there are 4 degenerate solutions, due to two well-known degeneracies. The first is
the usual close-wide degeneracy that occurs for events have planetary signals associated with the
central caustic. Normally, the close-wide degeneracy relates models that differ mainly in the 
$s \leftrightarrow 1/s$ substitution, but in this case with $s \sim 1$, the planetary caustics have
merged with the central caustic, which ruins the usual $s \leftrightarrow 1/s$ relation. So, the two
solutions have $s\approx 0.9$ and $s\approx 1.0$, instead.
The second degeneracy is the well known microlensing parallax degeneracy that involves
a flipping the orientation of the lens plane with respect to the orbit of the Earth. This is indicated
by sign changes of the $\alpha$ and $u_0$ parameters. 

One unexpected feature of this new analysis is that the best fit $\pi_{\rm E,N}$ value has changed
sign from the discovery paper \citep{koshimoto950} from $\sim 0.12$ to $\sim -0.22$, but in both
cases, the uncertainty is quite large, as is often the case for ground-based microlensing parallax
measurements \citep{muraki11,gould14}.
This change is due to an improvement in the detrending algorithm
that we have applied to the MOA data, and it does not make a significant difference in our final
conclusions.

In Section~\ref{sec-parallax}, we discuss constrained light curve models that employ constraints on the 
lens and source star relative proper motion, $\mubold_{\rm rel}$ and magnitudes that have been
derived in  Sections~\ref{sec-hst}, \ref{sec-Keck}, and \ref{sec-murel}. The best fit models
from this analysis are shown in Table~\ref{tab-Cmparams}. As we discuss in Section~\ref{sec-parallax},
the $\mubold_{\rm rel}$, the $\mubold_{\rm rel}$ constraints from our high angular resolution measuremetns
greatly improve the precision of our $\pi_{\rm E,N}$ measurements, which, in turn, enables precise
determinations of the masses of the host star and planet.

The final column in Tables~\ref{tab-UCmparams} and \ref{tab-Cmparams} is the average of each
parameter over Markov Chain Monte Carlo calculations for all the different models in a weighted sum. 
In Table~\ref{tab-UCmparams}, the weighting is based only on the $\chi^2$ difference ($\Delta\chi^2$)
between the different models, but in Table~\ref{tab-Cmparams}, we also include weightings from a
Galactic model. Note that the probability distributions the parameters, $u_0$, $s$, and $\alpha$, that take
significantly different values in the different degenerate models have double peaks.

\section{Follow up observations}
\label{sec-Followup}
The event OGLE-2012-BLG-0950 was observed with HST on May 22, 2018 with Wide Field Camera 3 - 
Ultraviolet Visible (WFC3-UVIS) instrument as part of the program GO 15455. Seven dithered images, each 
with 62 seconds exposure time, and eight dithered images, each with 111 seconds exposure time, 
were taken in F814W and F555W passbands (which are HST equivalent of $I$ and $V$ bands) respectively. 
The pixel scale for WFC3-UVIS instrument is $\sim$40 mas on a side. The images, corrected for CTE 
(Charge Transfer Efficiency) losses \citep{anderson2010},  were obtained from Mikulski Archive for Space 
Telescopes (MAST) and were reduced and stacked following the methods described 
in \citet{andking00,andking04}. The stars from the HST stack images were matched and calibrated to the
OGLE III catalog \citep{ogle3-phot}, which is already calibrated to Cousins $I$ \citep{cousins} and Johnson $V$ \citep{johnson}. 
Nine bright, isolated calibration stars with magnitude $I_{\rm OGLE_{III}} \leq 17.5$ and color 
$1.0 \leq (V-I)_{\rm OGLE_{III}} \leq 2.0$ were matched in both frames. We obtained the following calibration 
relations:
\begin{eqnarray}
I_{\rm OGLE_{III}} = 28.764 + I_{\rm HST} + 0.0467 (V-I)_{\rm HST} \pm 0.02\\
V_{\rm OGLE_{III}} = 30.602 + V_{\rm HST} - 0.0641 (V-I)_{\rm HST} \pm 0.03
\end{eqnarray}
These uncertainties are the RMS scatter of the fit, divided by the square root of the number of stars used for the transformation.

The same event was observed nearly simultaneously with the Keck AO (Adaptive Optics) NIRC2 
instrument during the  early 
morning of May 23, 2018 as part of our Keck NASA KSMS program. Five dithered exposures, 
each of 30 seconds, were taken in $K_S$ short passband with the wide camera. In this paper, from now 
on we refer to the $K_S$ band as the $K$ band. Each wide camera image covers a 1048 $\times$ 1048 square 
pixel area, and each pixel size is about $40 \times 40\,$mas$^2$. These images were flat field and dark current 
corrected using standard methods, and then stacked using the SWarp Astrometics package \citep{SWarp}. 
The details of our methods are described in \citet{batista2014}. We used aperture photometry method on these wide images with SExtractor software \citep{sextractor}. These wide images were used to detect 
and match as many bright isolated stars as possible to the VVV catalog for the calibration purposes. 
Twenty seven isolated bright stars in $K$ band were calibrated to VVV with a 0.02 magnitude dispersion.

This event was observed on the same date with Keck NIRC2 narrow camera in the $K$-band using laser guide star adaptive optics (LGSAO). The 
main purpose of these images is to derive a proper PSF for the astrometric and photometric analysis 
of the lens and source stars (section \ref{sec-Keck}). Thirty-nine dithered observations were taken with 
60 second exposures. The images were taken with a small dither of 0.7'' at a position angle (P.A.) of 0$^{\circ}$ with each frame consisting of 2 co-added 30 seconds integrations. The observations were taken in 8 dither positions with atleast 4 images in each dither position. The seeing of these narrow camera
images was $\sim$0.06-0.08''. There are 1048$\times$1048 pixels in each image with each pixel subtending
10 mas on each side. The stars from the narrow camera image were cross matched to the wide camera image for calibration. The photometry used for narrow camera images are from DAOPHOT analysis (section \ref{sec-Keck}).  

\section{ HST ePSF Fitting}
\label{sec-hst}
Like a many other space telescope cameras, the HST-WFC3-UVIS pixel scale undersamples the 
point spread function (PSF) in a compromise between field-of-view and angular resolution. Fortunately, 
accurate photometry \citep{lauer1999} and astrometry \citep{andking00} can still be obtained if the image 
pointings are dithered to recover the spatial sampling lost to undersampling. To overcome this problem, 
we adopt the method of \citet{andking06} to construct an Effective Point Spread Function (ePSF) from 
the dithered images. This method has proved effective at measuring the separation of stars separated
by $< 1$ Full-Width-Half-Maximum (FWHM) \citep{ogle169,moa310}.
Eight main sequence stars, within 120 pixel radius of the target and similar 
brightness as the target were chosen to build the ePSF. The ePSFs of all these eight stars were 
computed in each image frame and then the average ePSF over all the frames was obtained. It is this 
averaging of the ePSFs from all dithered images that helps to overcome the undersampling problem. 
The procedure was iterated until the average ePSFs converged. We used this final ePSF model to fit single 
and dual stellar profiles for our target object as described in next paragraphs. Our methodology for fitting 
single and multiple stellar profiles with ePSFs is described in detail in \citet{moa310}. 
  
\begin{figure}
\epsscale{1.0}
\plotone{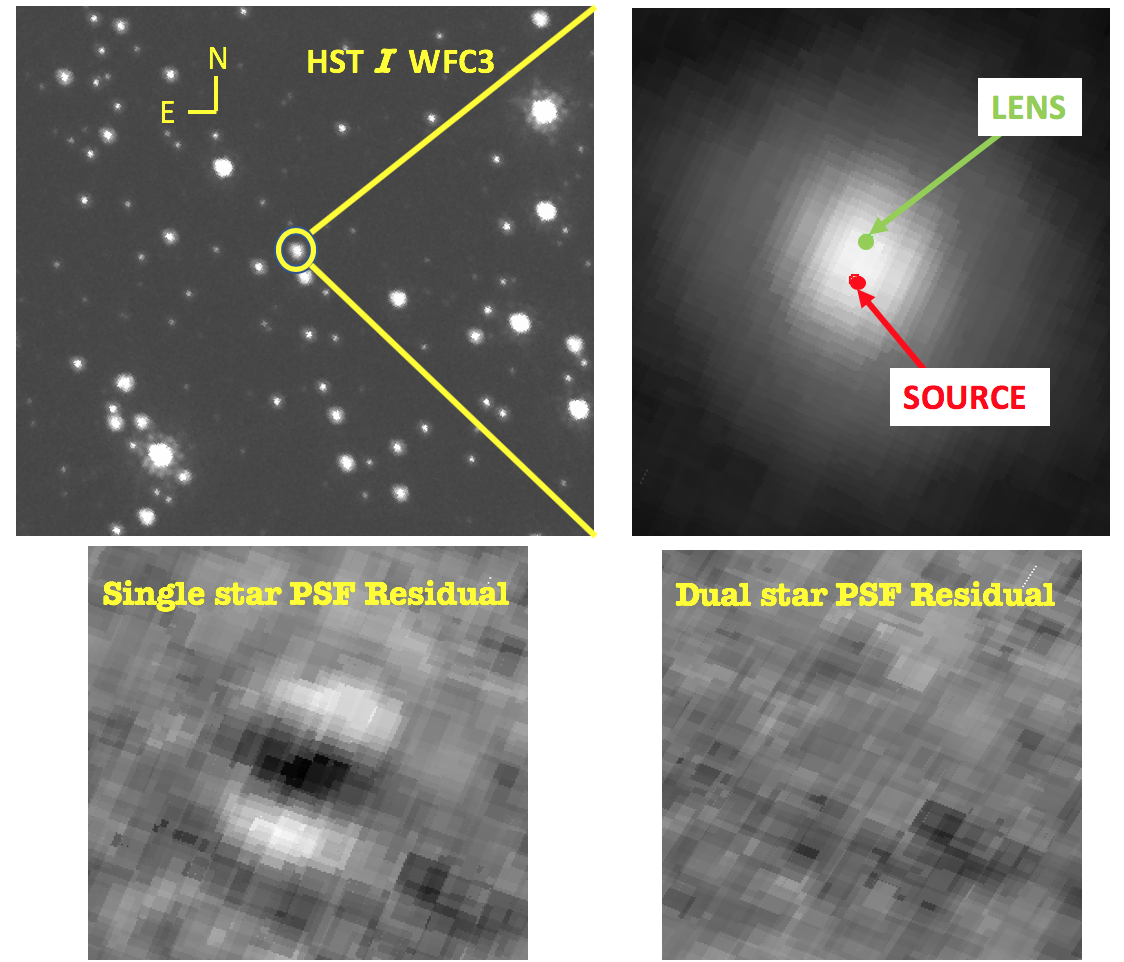}
\caption{{\it Top left}: The stack image in the HST $I$ (F814W) passband with the target indicated by 
the yellow circle. {\it Top right}: The $100\times$ supersampled summed image of the target object. 
The source and lens positions obtained from the best fit dual star PSF model to the individual images. 
{\it Bottom left}: The residual image from the best fit single star PSF model. The wings are under-subtracted
and the core is over-subtracted indicating that the best fit single star PSF model is not consistent with the 
flux distribution of the target object. {\it Bottom right}: The residual image from the best fit dual star 
PSF model. This indicates a much better fit consistent with Poisson noise (which is larger at the location
of the subtracted stars), so the best fit dual star model is compatible with the flux distribution of the target. Both the bottom images are demonstrated using the same photometry scale.}
\label{fig-hst}
\end{figure}
  
The first step of ePSF modeling is to do a single star fit of our target object. There are three model 
parameters for such a fit: the two dimensional position of the star and the stellar flux. We selected a 
region centered on the target that included 151 pixels in our 7 $I$-band images and 200 pixels in our 8 
$V$-band images. The calibrated magnitudes of the target object from the best fit solution were 
$18.64 \pm 0.02$ and $20.41 \pm 0.03$ in the $I$ and $V$ bands, respectively. The magnitude uncertainties 
are the combination of the ePSF fitting and calibration uncertainties. As Figure \ref{fig-hst} shows, the 
residual image from the fit indicates that the best single star fit was a poor fit. It was also clear that the $I$ 
and $V$ magnitudes of the target were significantly brighter than the source magnitude determined from 
the microlensing light curve analysis: $I_S = 19.26 \pm 0.05$, and $V_S= 20.65 \pm 0.07$. 
This indicates the presence of at least one additional star blended with the source star. This leads to
the dual star ePSF fitting analyses described below.    

\begin{deluxetable}{ccc}
\tablecaption{Photometry from Dual Star PSF Fits \label{tab-dual-fit}}
\tablewidth{0pt}
\tablehead{\colhead{Star}&\colhead{Passband}& Mag}
\startdata
Lens  &HST $I$& $19.57\pm 0.09$\\
      &HST $V$& $22.27\pm 0.21$\\ 
      &Keck $K$& $17.27\pm 0.04$\\
Source &HST $I$& $19.24\pm 0.06$\\
      &HST $V$& $20.65\pm 0.09$\\ 
      &Keck $K$& $17.68\pm 0.05$\\
\enddata
\\$^*$Magnitudes are calibrated Cousins $I$-band, Johnson\\
 $V$-band and 2MASS $K$-band magnitudes.
\end{deluxetable}
  
The dual star ePSF fitting method requires six model parameters: the two dimensional positions of each 
of the stars ({$x_1,y_1$},{$x_2,y_2$}), the total flux ($Z$) and the flux fraction ($f_1$) of star \# 1. 
The first stage or our dual star fitting process is a grid search. The positions of each star were allowed to 
explore the full grid of 151 and 200 pixels for $I$ and $V$ bands with a step size of 0.02 pixel. 
For each combination of positions of the two stars, the flux fraction, $f_1$, was varied between (0.0 -1.0) 
with a step size of 0.02. At each step, the $\chi^2$ value was computed and the minimum $\chi^2$ 
solution was stored. This intense grid search yielded our first best fit result on the dual star ePSF fit. 
The grid search was performed to cover the full parameter space. The best fit result was used as an 
initial condition for a  Markov Chain Monte Carlo (MCMC) \citep{mcmc} ePSF fitting in order to obtain
a more precise measurement with error bars. Our detailed methodology of MCMC ePSF fitting for multiple 
stars is described in detail in \citet{moa310}. The best fit photometry from the dual star MCMC PSF models is 
shown in Table \ref{tab-dual-fit}, and the best fit relative astrometry is shown in Figure~\ref{tab-sep}. 
The uncertainties are determined from the distribution of 29727 and 37132 accepted MCMC links in the
$I$ and $V$-bands, respectively. The residual of the dual star shown in Figure \ref{fig-hst} shows that it is a 
good model.   
  
\begin{deluxetable}{ccccc}
\tablecaption{Measured Lens-Source Separation and Relative Proper Motion\label{tab-sep}}
\tablewidth{0pt}
\tablehead{\colhead{Passband}&\multicolumn{2}{c}{Separation(mas)}&\multicolumn{2}{c}{$\mathbf{\mu}_{\rm rel,H}$(mas/yr)}\\
&East&North&East&North}
\startdata
HST $I$ & $-15.5\pm 0.4$ & $30.5\pm 0.7$ & $-2.66\pm 0.05$ &$5.23\pm 0.13$\\
HST $V$ & $-15.6\pm 1.4$ & $30.8\pm 2.6$ & $-2.68\pm 0.27$ & $5.28\pm 0.47$\\
Keck $K$& $-15.9\pm 2.9$  & $29.3\pm 3.0$& $-2.7\pm 0.49$ & $5.03\pm 0.52$\\
\enddata
\end{deluxetable}

The dual star ePSF model results (Table \ref{tab-sep}) show that the target consists of two stars with
a separation of $\sim 34\,$mas, with consistent separations measured in all three passbands.
The magnitudes of star \# 1 (the southernmost star) are $I_1 = 19.24 \pm 0.06$ and $V_1 = 20.65 \pm 0.09$, and 
the magnitudes of the second star are $I_2 = 19.57 \pm 0.09$ and $V_2 = 22.27 \pm 0.21$. The magnitudes for star \# 1 match the approximately calibrated source magnitudes \citep{koshimoto950}, 
and they are within 1$\sigma$ of the source brightness predicted from our light curve modeling based on
improved MOA photometry. However, star \# 2 does not match with the predicted source flux from light curve of $I_S = 19.24 \pm 0.03$ and $V_S = 20.65 \pm 0.07$. Thus, 
we identify star \# 1 as the source star. Since both stars have similar brightnesses in the $I$ band and 
the lens is located nearer than the source in the bulge, the lens should be redder than the main 
sequence source. So, the star we identify as the source is the bluest and significantly brighter in the $V$ 
band. The second star is a candidate for the planetary host and lens star. The lens-source separations in 
both East and North are consistent in independent analysis of the $I$ and the $V$ passbands. 
The separation of these two stars is also measured to high precision. The one dimensional separation,
as measured in the $I$-band is $34.2 \pm 0.06\,$mas, which is a precision of 2\%, even though the
separation is $< 0.5\,$FWHM of the PSF. This is consistent with our previous analysis of the event
MOA-2008-BLG-310, where we were able to measure the separation between the source and
blend star (which was not the lens) to a precision of 27\%, when the separation was $12\,$mas
\citep{moa310}.

\section{Keck AO PSF Fitting}
\label{sec-Keck}    

Thirty-nine images taken with the Keck NIRC2 narrow camera were reduced. These images
taken with the narrow camera are not undersampled, so we did not need to adopt the ePSF method for 
the analysis of these images. For the reduction of these images, we used $K$-band dome flats 
taken with narrow camera on the same day as the science images. There were 5 dome flat  images with 
the lamp on and 5 more images with the lamp off, each with 65 seconds exposure time. Also at the end of 
the night, we took 20 sky images using a clear patch of sky at a (RA, Dec) of (18:08:04.62 , -29:43:53.7) 
with an exposure time  of 30 seconds each. All these images were used to flat field, bias subtract and 
remove bad pixels and cosmic rays from the thirty-nine raw science images. Finally 
these clean raw images were stacked into one image and that we used for the final photometry and
astrometry analysis. 

Because the source and candidate lens stars are separated by $< 1\,$FWHM, we must analyze the Keck
data with a PSF fitting code to measure the astrometry and photometry of this 2-star system. However,
PSF fitting photometry codes that employ analytic PSF models may have problems with AO images that
often have highly non-Gaussian PSFs. We have previously found that DAOPHOT \citep{Daophot} does 
quite well with such images  \citep{bennett2010}, so we use DAOPHOT for the analysis of our
OGLE-2012-BLG-0950 Keck data.

To start our analysis, we needed to construct a proper PSF. We built the PSF in three stages. In the first 
stage, we run the FIND and PHOTOMETRY commands of DAOPHOT to find all the possible stars in the 
image. Then we used the PICK command to find twenty-three bright ($K < 19.24$) isolated stars to be 
used for constructing our PSF. Our target object was excluded from this list of PSF stars because it
is expected to consist of two stars that are not in the same position. We build a PSF from these stars and 
fit all the stars in the field with this PSF. In the second stage, we carefully check the residual image that has
all the identified stars subtracted. We noticed several stars which had large residuals. These were
either very bright stars that were near saturation (where the detector becomes non-linear) or elongated
PSFs due to multiple blended stellar images. At this stage, we carefully checked and found that two of 
our PSF contributing stars had a significant residual from the PSF subtraction. A close look showed that both of
these had slight elongations that are probably due to binary companions. So, these two stars were removed
from our PSF star list, and the PSF was constructed again. Next, we ran the PSF fitting again on the field 
with the new PSF and moved on to the third stage. In the third stage, we removed all the neighbor stars 
of the PSF stars and computed a clean final PSF from these 21 stars. We then did a final round of PSF fitting
for all the stars in the image with this clean PSF. 

\begin{figure}
\epsscale{1.0}
\plotone{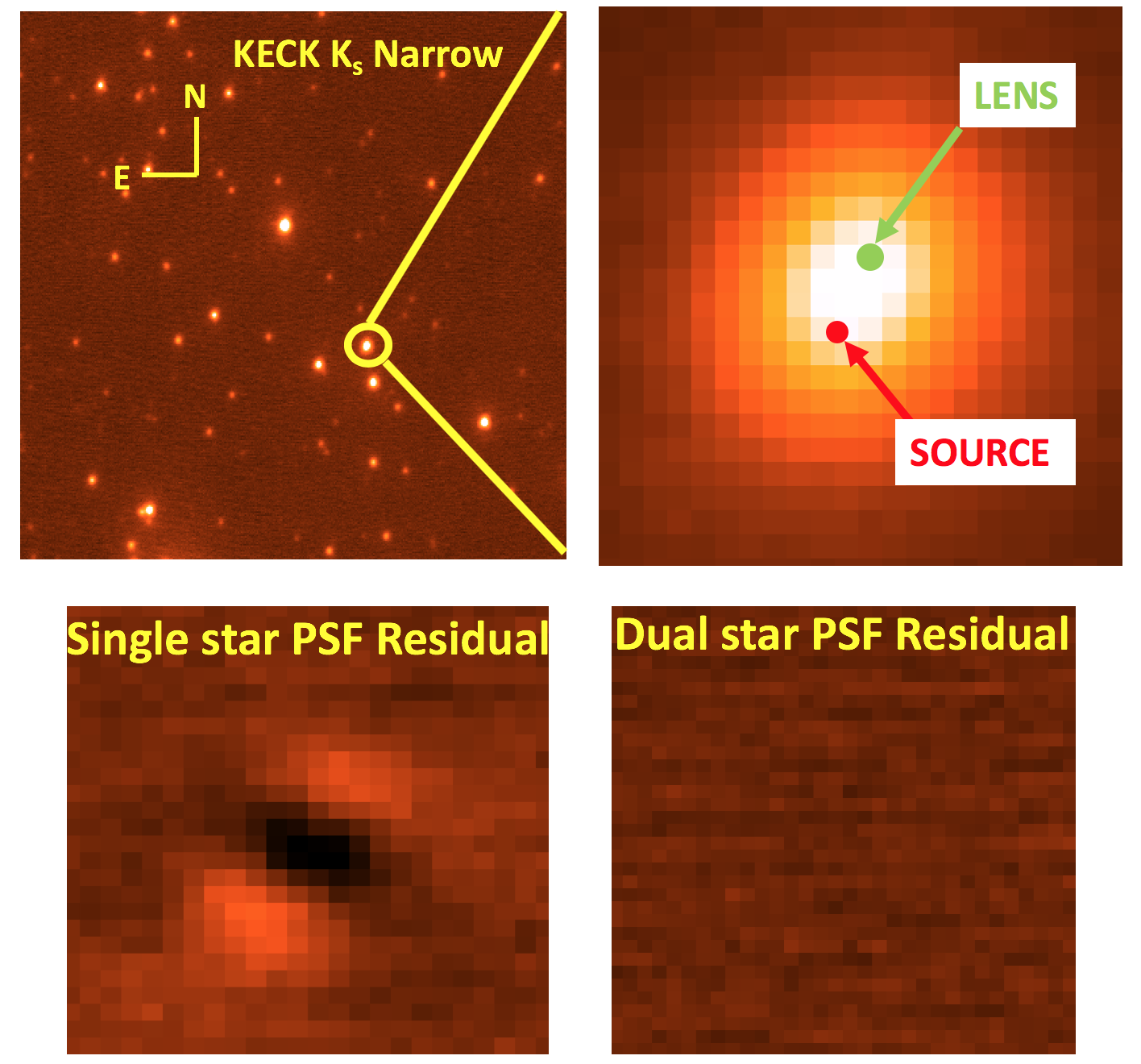}
\caption{{\it Top left}: The stack image of 39 Keck $K$-band images, taken with the narrow camera. The 
target is indicated by the yellow circle. {\it Top right}: A closer look at the target object. The source and lens 
positions are obtained from the best fit dual star PSF model. {\it Bottom left}: The residual image after subtracting 
the best fit single star PSF model. The under-subtracted wings and the over-subtracted core indicates that 
the best fit single star PSF model is not consistent with the data. {\it Bottom right}: The residual image after 
subtracting the best fit dual star PSF model. This shows only noise, which implies that the best fit dual star 
model can account for the flux distribution of the target. Both the bottom images are demonstrated using the same photometry scale.}
\label{fig-keck}
\end{figure}

After finding a good PSF model, we started our analysis with a single star PSF fit to the target object. 
The residual of this fit is shown in Figure \ref{fig-keck}. This residual shows a clear pattern that indicates
that it is elongated compared to the PSFs of single stars. From our reanalysis of this event 
(with improved MOA photometry), we find an extinction corrected magnitude and color of the
source star $I_{S,0} = 18.40  \pm 0.07$ and $(V-I)_{S,0}=0.74\pm 0.07$, which is consistent with the
\citet{koshimoto950} analysis. From the color-color relation in \citet{kenyon_hartmann}, 
$(V-I)_{S,0}=0.74$ corresponds to the dereddened color $(I-K)_{S,0}=0.75$. From \citet{cardelli}, the extinction 
in $K$ band is $A_K = 0.12$ at 8.2 kpc (see section \ref{sec-lens}). Hence, the source $K$ band 
calibrated magnitude is $(18.40 -0.75 +0.12) \pm 0.07$ = $17.77 \pm 0.07$. 

The total measured $K$-band
flux from our single star fit was $16.71 \pm 0.03$ which indicated an excess flux of $K = 17.22$
on top of the source. This excess flux, combined with the residual shown in the lower left panel of
Figure~\ref{fig-keck} implies that we should proceed with dual star PSF fitting for this target.
The dual star fit with DAOPHOT limited the stars to move up to a 30 mas or 3 pixel radius from their initial positions. 
In the HST astrometry analysis (section \ref{fig-hst}), we found that the separation between the two stars is 
$\sim$ 34 mas. So,  the movement of star positions by a limit of 30 pixels should not impede our
dual star modeling. The best dual star fit yielded two stars with calibrated $K$ magnitude $17.27 \pm 0.04$ 
and $17.68 \pm 0.05$, as listed in Table~\ref{tab-dual-fit}, which matches the
predicted source brightness from the discovery paper. The separation of these two stars, given in 
Table~\ref{tab-sep} is consistent with their separations in the HST $I$ and $V$ bands. Both the single star and dual star fits were done using Newton-Raphson method following \citet{Daophot}. The uncertainties were calculated following \citet{king1983}: 
\begin{equation}
\sigma_{x} = 0.65238 \times \rm{FWHM} \times \sqrt{\frac{4}{3}}\times \frac{\sigma_{ \rm{F}}}{\rm {F}} 
\end{equation}
 where F is the flux of the respective star. The uncertainty in y direction can be presented by the same equation. The FWHM of the K band narrow image measured from DAOPHOT is 84 mas and 87 mas in x and y directions. 
  
\section{Color Dependent Centroid Shift with Simultaneous HST and Keck Observations}
\label{sec-centroid}

\begin{figure}
\epsscale{1.0}
\plotone{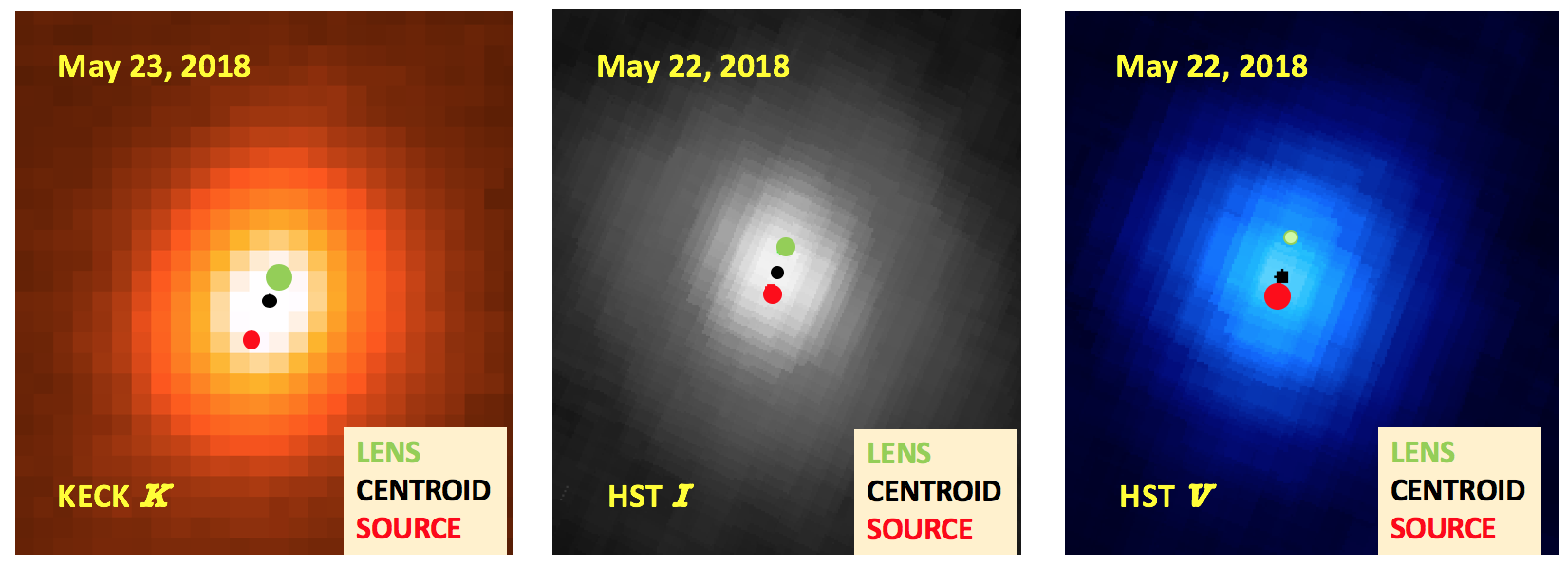}
\caption{The blended image of the source plus lens stars is shown in three different passbands: 
Keck $K$ (left), HST $I$ (middle) and HST $V$ (right). The source and lens positions, determined 
from the best fit dual star PSF models, are shown in red and green dots, respectively. The centroids 
are calculated according to the flux of the lens and the source in the respective passbands (from 
Table \ref{tab-dual-fit}). The size of the source and lens dots are proportional to the flux fraction for the lens 
and source stars. In the Keck $K$-band the lens is brighter than the source hence the green dot is bigger 
than the red dot and the centroid is shifted toward the lens. In the $I$-band, the source is slightly brighter
than the lens, so the centroid is slightly closer to the source. In the $V$-band, the centroid is moved toward the 
source, since the source is much brighter than the lens.}
\label{fig-centroid}
\end{figure}

The color dependent centroid shift is a method that can be used to confirm the identification of excess
blended flux on top of the source with the lens star. This is possible because of the constraints on the
properties of the blended image of the lens and source are known from the microlensing light curve.
These known properties always include the source star brightness, and they usually include the lens-source 
relative proper motion, $\mu_{\rm rel}$ or the microlensing parallax, $\pi_E$. 

The source stars are biased toward the brightest stars in the bulge, since a brighter source provides a stronger microlensing signal. However, the lens stars are detected with a probability
that scales as lens mass, $M_{L}^n$ where $0.5 \leq n \leq 1$, which is a much shallower dependence. Thus, the lens stars 
tend to have a lower mass with a redder color than the source stars as the cases of OGLE-2005-BLG-169
\citep{ogle169,batistaogle169} and OGLE-2012-BLG-0950, presented in this paper, demonstrate.

When the source and lens stars have different colors, their blended image will have different centroids
in follow-up images in different passbands due to the separation between the lens and source. This is
the color dependent centroid shift, which was first demonstrated for the first planet found by microlensing
\citep{bennett06} using HST data. The main advantage of this method over the measurement of 
image elongation is that the S/N of color dependent centroid shift scales as time interval since the 
magnification peak, $\Delta t$, whereas the image elongation signal grows as $\Delta t^2$
if the source brightness is known, or as $\Delta t^3$ when the source
brightness is not constrained \citep{bennett07}.
Thus, the color dependent centroid method might be the most effective method for determining the
lens-source separation shortly after an event.

\begin{deluxetable}{lrrrrrr}
\tablecaption{Blended Lens + Source Centroid Shifts \label{tab-col-centroid}}
\tablewidth{0pt}
\tablehead{
& \multicolumn{2}{c} {Single-Star} & \multicolumn{2}{c} {Dual-Star} & \multicolumn{2}{c} {Single-Star -- Dual-Star}  \\
\colhead{Passbands}& $\Delta{\rm E}$(mas) & $\Delta{\rm N}$(mas) & $\Delta{\rm E}$(mas) & $\Delta{\rm N}$(mas) & $\Delta{\rm E}$(mas) & $\Delta{\rm N}$(mas) }
\startdata
$V-I$      & $-3.764$ & $7.249$ & $-3.460$ & $5.950$ & $-0.304$ & $1.299$ \\ 
$V-K_S$& $-6.400$ & $12.324$ & $-6.815$ & $13.358$ & $0.415$ & $-1.034$ \\ 
$I-K_S$ & $-2.635$ & $5.074$ & $-5.709$ & $7.863$ & $3.074$ & $-2.789$ \\ 
\enddata
\\$^*$The positions are relative to the $K$-band position of the source at\\
 RA 18:08:04.61228 and Dec -29:43:53.3565 (J2000).
\end{deluxetable}

The color dependent centroid shift is largest with the maximum color difference between passbands, and
this was the justification for the near simultaneous HST and Keck AO imaging. Previous studies \citep{lu14}
have argued that it is possible to obtain astrometry with an accuracy of $150\,$mas from Keck NIRC2 images,
but this requires a large number of careful corrections. Since our primary science result comes from the
dual star fits, we do not require this very high precision astrometry for our science results. So, we present
a preliminary test of the color dependent centroid method here. We plan to present more a more refined
analysis in a future paper.

In order to measure the color dependent centroid shift between each pair of passbands, we must
perform a coordinate transformation between the images in the different passbands. We have corrected
for achromatic differential refraction \citep{refraction} and geometric distortion \citep{distortion}. Based on \citet{gubler}, the chromatic differential refraction effect would be < 0.2 mas. Hence this effect is ignored for our preliminary analysis. For our preliminary analysis, we have performed linear coordinate transformations
with a relatively small number of stars around the target.
For the HST-$I$ to Keck-$K$ transformation, we used 10 stars, which resulted in a one-dimensional 
RMS scatter of $0.9\,$mas. For the HST-$V$ to Keck-$K$ transformation, we used 8 stars with a 
one-dimensional RMS scatter of $1.3\,$mas. Finally, for the HST-$I$ to HST-$V$ transformation, we
used 19 stars with an RMS scatter of $0.2\,$mas, which is consistent with the astrometric precision
obtained in a previous attempt to measure the color dependent centroid shift \citep{bennett06}.
These coordinate transformations were used to compare the centroids of the blended lens plus
source image from the single PSF fits, and these results are reported in the second and third columns
of Table~\ref{tab-col-centroid}.

We can also use our dual star fits from Sections~\ref{sec-hst} and \ref{sec-Keck} to calculate the
expected color dependent centroid shifts. From section \ref{sec-hst}, the flux fraction of the source 
(or star \# 1) , is $f_1$, which implies that the flux fraction of the excess flux is $(1-f_1)$. For the remainder
of this section we will refer to this excess flux as being due to the lens star.
(see Section \ref{sec-lens} and Figure \ref{fig-mass}). We denote the source position vector as $\bm{x}_1$ and the lens 
position as $\bm{x}_2$. The lens -source separation is given by $\Delta \bm{x}_{L-S}$.  
For a passband ``i" the centroid of the combined source and the lens flux, $\bm{x}_{c,i}$ is given by:  
\begin{gather}
\bm{x}_{c,i} = f_{1,i}\bm{x}_1 + (1-f_{1,i})\bm{x}_2\\
\bm{x}_2 = \bm{x}_1 +\Delta \bm{x}_{L-S}\\
\bm{x}_{c,i} = f_{1,i}\bm{x}_1 + (1-f_{1,i})(\bm{x}_1 +\Delta \bm{x}_{L-S})
\end{gather}
For a different passband $j \neq i$, the centroid of the blended source plus lens image is given by:
\begin{gather}
\bm{x}_{c,j} = f_{1,j}\bm{x}_1 + (1-f_{1,j})(\bm{x}_1 +\Delta \bm{x}_{L-S})
\end{gather}
Subtracting Equation (6) from Equation (7) we obtain the centroid shift between passbands $j$ and 
$i$ as shown in Equation (8). Rearranging the terms, we can derive the lens-source separation from 
this centroid shift and the flux ratios in two different passbands.  
\begin{gather}
\Delta \bm{x}_{c,(j,i)}= \bm{x}_1(f_{1,j}-f_{1,i})+(f_{1,i}-f_{1,j})(\bm{x}_1 + \Delta \bm{x}_{L-S})\\
\Delta \bm{x}_{c,(j,i)}=(f_{1,i}-f_{1,j})\Delta \bm{x}_{L-S}\\
\Delta \bm{x}_{L-S} = \frac{\Delta \bm{x}_{c,(j,i)}}{(f_{1,i}-f_{1,j})}
\end{gather}

We can now use Equation (9) to predict the color dependent centroid shifts from the dual star fits. The
results are shown in the fourth and fifth columns of Table~\ref{tab-col-centroid}. The sixth and seventh
columns of this table show the difference between these two predictions. The centroid shift estimate from the 
dual star fits seems to be a rough match to the centroid measurements from the single star fits, but the
difference between the single star fit measurements and the dual star fit estimates are much larger
than we would like if we are to use the color dependent centroid shift measurement to confirm lens-source
separation predictions at small separations. We expect that an astrometric precision in each of the 
passbands of $\sim 0.3\,$mas to be achievable \citep{bennett06,lu14}, so that we should be able to achieve 
relative astrometric precision of $\simlt 0.5\,$mas between pairs of passbands.  However, based on the scatter
in our transformations between passbands, it is clear that this is only possible between the HST $V$ and 
$I$-bands. Our transformations between the HST passbands and the Keck $K$-band have a one-dimensional
scatter of $\simgt 1\,$mas, so sub-mas precision is not possible until these transformation are improved.
Fortunately, there are several ways in which these transformation can be improved. First, we need to double-check
the differential refraction and optical distortion corrections that we have made to the $K$-band data. \citep{lu14} have handled this with higher order coordinate transformations. This could be a 
problem with only 8-10 comparison stars, however. We will try increasing the number of comparison stars
by selecting stars closer to the edge of the Keck narrow frames, but we will also consider changes in our
observing program for future simultaneous Keck plus HST observing campaigns. For example, we might take
deeper NIRC2-narrow camera images to allow the use of fainter comparison stars. Or we might try the
NIRC2-medium camera to increase the field-of-view by a factor of 4, but with the pixel size increased from
10 to $20\,$mas.

We should note that the lens-source separation of $34\,$mas could be too large for Equations
4-10 to apply, particularly in the HST $V$-band with a PSF FWHM of $\sim 48\,$mas. Thus, the 
comparison shown in Table~\ref{tab-col-centroid} might not be completely fair. From the Table 3, we can conclude that the positions of the lens derived from the single star fit would be different by $\sim 10-20 \%$ from the measured positions of the lens using dual fits. Hence, the large elongation could be a big reason why this centroid shift method is not yielding a result close to the dual fits. 

\section{Determination of Relative Lens-Source Proper Motion}
\label{sec-murel}

Our high resolution observations were taken 5.83 years after the microlensing event magnification peak. If these images were taken 6 years after the microlensing magnification, then the HST frame would have coincided with the heliocentric frame. However, this time difference of 0.17 years between the heliocentric frame and our current HST frame, with a lens system at 2.1 kpc (see section \ref{sec-lens})  produces less than 1$\sigma$ difference in the lens-source separation. Hence we approximate the HST frame to the heliocentric frame. At the time of peak magnification, the separation between lens and source was $\sim u_0\theta_E \sim 0.1$ mas. Hence, by dividing the measured separation by the time interval 5.83 years, we obtained the heliocentric lens-source 
relative proper motion, $\mubold_{\rm rel,H}$. A comparison of these values from our independent dual star
fits is shown in Table \ref{tab-sep}. In the galactic coordinates ,the $\mubold_{\rm rel,H}$ value 
from the $I$-band measurement is $\mu_{{\rm rel,H},l} = 3.23\,$mas/yr and  
$\mu_{{\rm rel,H},b} = 4.84\,$mas/yr, with an amplitude
of $\mu_{\rm rel,H} = 5.87\pm 0.12$ at an angle of $\sim 57^\circ$ from the direction of Galactic rotation. 
The dispersion in the motion of stars in the local Galactic disk is $\sim 30\,$km/sec, which corresponds
to a proper motion dispersion of $\sim 3\,$mas/yr (in both directions) at the lens distance of 
$D_L \approx 2.1\,$kpc as presented in section \ref{sec-lens}.  The source is in the bulge, with a proper 
motion dispersion of $\sim 2.5\,$mas/yr in each direction. Thus, the measured $\mubold_{\rm rel,H}$ is
entirely consistent with the combination of the mean relative proper motion of $6\,$mas/yr in the 
direction of Galactic rotation combined with the proper motion dispersion of bulge source and disk
lens stars.

Our light curve models were done in a geocentric reference frame that differs from the heliocentric
frame by the instantaneous velocity of the Earth at the time of peak magnification, because the 
light curve parameters can be determined most precisely in this frame. However, this also means that
the lens-source relative proper motion that we measure with follow-up observations is not in the
same reference frame as the light curve parameters. This is an important issue because, as we
show below, the measured relative proper motion can be combined with the microlensing parallax 
light curve parameter to determine the mass of the lens system. The relation between the relative
proper motions in the heliocentric and geocentric coordinate systems are given by \citep{dong-moa400}:   
\begin{equation}
\bm{\mu}_{\rm rel,H} = \bm{\mu}_{\rm rel,G} + \frac{{\bm v}_{\oplus} \pi_{\rm rel}}{\rm AU}  \ ,
\label{eq-mu_helio}
\end{equation}
where ${\bm v}_{\oplus}$ is the projected velocity of the earth relative to the sun (perpendicular to the 
line-of-sight) at the time of peak magnification. The projected velocity for OGLE-2012-BLG-0950 is
${{\bm v}_{\oplus}}_{\rm E, N}$ = (4.096, -0.448) AU/yr at the peak of the microlensing, HJD'= 6151.48. The relative parallax is 
defined as $\pi_{\rm rel} \equiv 1/D_L - 1/D_S$, where $D_L$ and $D_S$ are lens and source distances. 
The lens distance for this event can be determined in two different ways. At each possible lens distance, 
we can use the $\mu_{\rm rel,G}$ value from equation~\ref{eq-mu_helio} to determine the angular 
Einstein radius, $\theta_E = \mu_{\rm rel,G} t_E$. As we explain below, the $\bm{\mu}_{\rm rel,G}$
can also be used to convert a one-dimensional microlensing parallax measurement into a full
measurement of the microlensing parallax vector. The angular Einstein radius, microlensing parallax,
and the three lens flux measurements in the $V$, $I$, and $K$-bands all constrain the mass and distance
of the lens, as we explain in the next section.

\section{$\bm{\mu}_{\rm rel,H}$ and Lens Flux Constraints on ${\bm{\pi}_{\rm E}}$ and Light Curve Models}
\label{sec-parallax}

\begin{figure}
\epsscale{0.9}
\plotone{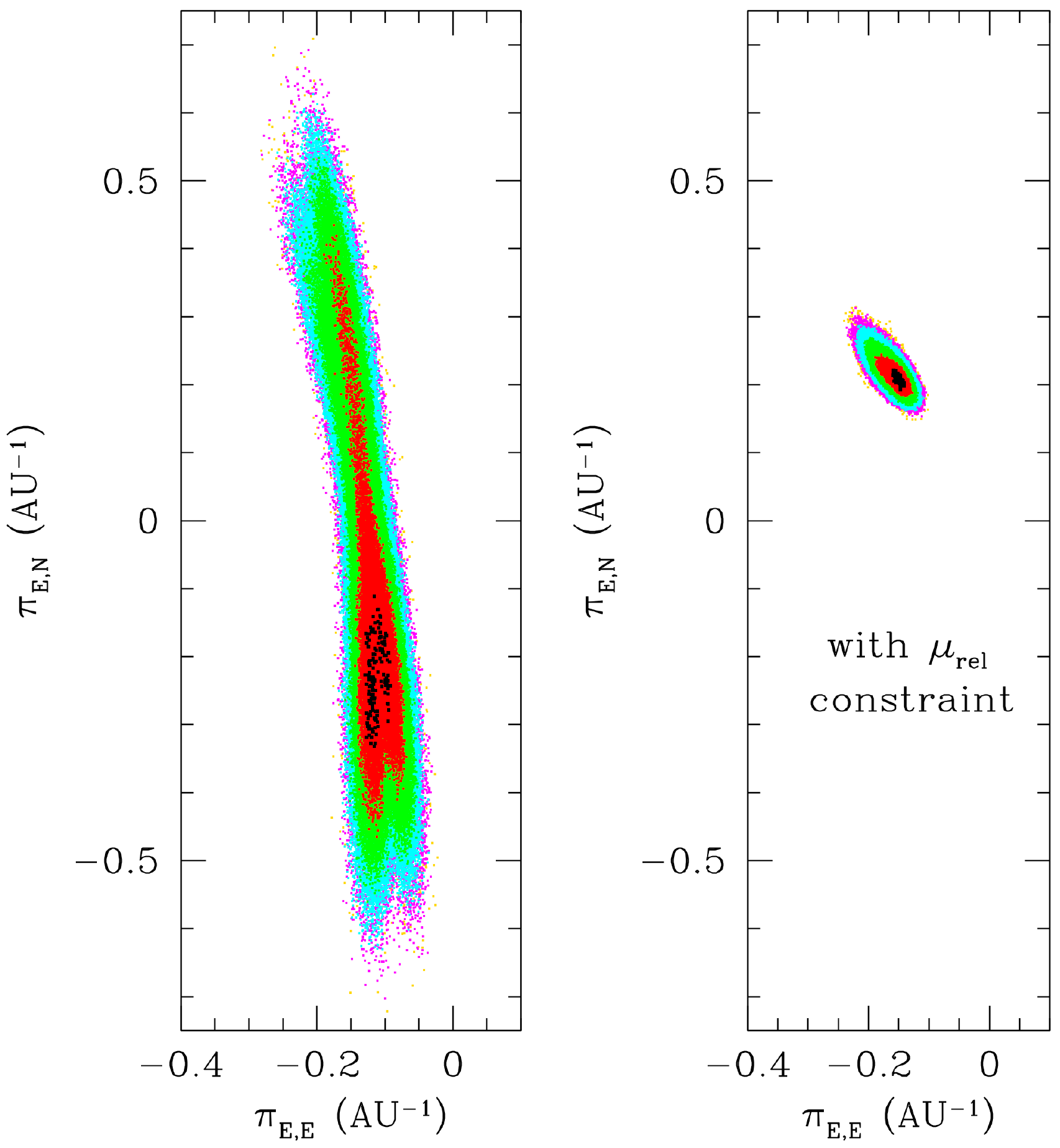}
\caption{{\it Left panel}: The ${\bm \pi_{\rm E}}$ distribution from light curve modeling without any constraint
from follow-up observations.
{\it Right panel}: The ${\bm \pi_{\rm E}}$ distribution resulting from the addition of the high resolution follow-up
imaging constraints. The following color scheme is used to denote the $\chi^2$ differences from the best fit 
light curve model: black represents $\Delta\chi^2 < 1$, red represents $\Delta\chi^2 < 4$, 
green represents $\Delta\chi^2 < 16$, cyan represents $\Delta\chi^2 < 25$, and magenta represents 
$\Delta\chi^2 \geq 25$. The right panel clearly shows that the relative proper motion measurements from 
HST and Keck constrain $\pi_{\rm E, N}$, which is  the North component of ${\bm \pi_{\rm E}}$, that 
was largely unconstrained by the light curve. Without the $\bm{\mu}_{\rm rel,H}$ measurement, in the left
panel, the light curve slightly favors solutions with $\pi_{\rm E, N} < 0$, but the constraint forces 
$\pi_{\rm E, N} > 0$. Note that this figure combines both the degenerate $u_0 > 0$ and $u_0 < 0$ models.}
\label{fig-parallax}
\end{figure}

The OGLE-2012-BLG-0950 light curve shows a significant improvement of $\Delta\chi^2 = 85.9$ due to 
the measurement of the microlensing parallax effect. But as it is often the case
\citep{muraki11,gould14}, only the $\pi_{E,E}$ 
component of the microlensing parallax vector is measured precisely. As shown in the left panel of 
Figure~\ref{fig-parallax}, the 2-$\sigma$ range for $\pi_{E,N}$ is $-0.39 < \pi_{E,N} < 0.43$. However,
the microlensing parallax vector, ${\bm{\pi}_{\rm E}}$ is parallel to the $\bm{\mu}_{\rm rel,G}$ vector, and 
the two quantities are related by
\begin{equation}
{\bm{\pi}_{\rm E}} = \frac{\pi_{\rm rel}}{t_{\rm E}}\frac{\bm{\mu}_{\rm rel,G}}{|\mu_{\rm rel,G}|^2} \ ,
\label{eq-piE_muG}
\end{equation}
so with measurements of $\pi_{E,E}$ and $\bm{\mu}_{\rm rel,H}$, we can use equations~\ref{eq-mu_helio}
and \ref{eq-piE_muG} to solve for $\pi_{E,N}$ \citep{ghosh04,bennett07}. \citet{gould14} has shown that
equations~\ref{eq-mu_helio} and \ref{eq-piE_muG} can be converted to a quadratic equation in $\pi_{E,N}$.
A quadratic equation means two solutions, and Gould argued that this presented an important degeneracy
that could lead to an ambiguous interpretation, but we find that this is generally not the case. For our measured 
value of $\bm{\mu}_{\rm rel,H}$, the degenerate solutions require that one solution had $\pi_{\rm rel} < 0$, 
which would imply that the lens is (unphysically) more distant than the source. So, there is a unique 
solution in the case of OGLE-2012-BLG-0950. If the sign of $\bm{\mu}_{\rm rel,H}$ was reversed, then 
there would be some degeneracy in the $\pi_{E,E}$ values at large $|\pi_{E,E}|$, but these would only
be important for $D_L < 0.08\,$kpc. In general, this degeneracy is not important when 
$|\bm{\mu}_{\rm rel,H}| \gg | {\bm v}_{\oplus} \pi_{\rm rel}/{\rm AU}|$, as is the case for virtually all 
microlensing events observed towards the Galactic bulge.

In order to obtain good sampling of light curves that are consistent with our constraints, we apply constraints
inside our modeling code to ensure that the Heliocentric proper motion and lens magnitudes are 
consistent with the Keck and HST observations. These constraints are $\mu_{\rm rel,H,N} = 5.33 \pm 0.26$,  
$\mu_{\rm rel,H,E} = -2.46 \pm 0.26$, $V_L = 22.18 \pm 0.26$, $I_L = 19.51 \pm 0.09$, and 
$K_L = 17.21 \pm 0.14$. They are implemented by calculating a $\chi^2$ contribution from each of the 
constraints and adding it to the light curve fit $\chi^2$ inside the modeling code \citep{bennett-himag}.
This requires the use of a mass-luminosity relation. As argued in \citet{bennett_moa291}, an empirical
mass luminosity relation is preferred for lens masses $\lesssim 0.7\msun$. Following \citet{bennett_moa291},
we use a combination of mass-luminosity relations for different masses. 
For $M_L \geq 0.66\,\msun$, $0.54\,\msun\geq M_L \geq 0.12\,\msun $, and 
$0.10 \,\msun \geq M_L \geq 0.07\,\msun$, we use the relations of \citet{henry93}, \citet{delfosse00},
and \citet{henry99}, respectively. In between these
mass ranges, we linearly interpolate between the two relations used on the
boundaries. That is, we interpolate between the \citet{henry93} and the \citet{delfosse00}
relations for $0.66\,\msun > M_L > 0.54\,\msun$, and we interpolate between the
\citet{delfosse00} and \citet{henry99} relations for $0.12\,\msun > M_L > 0.10\,\msun$.

For the mass-luminosity relations, we must also consider the foreground extinction.
At a Galactic latitude of $ b = -4.634^\circ$, and a lens distance of $\sim 2\,$kpc, the lens system
is likely to be behind some, but not all, of the dust that is in the foreground of the source. 
We assume a dust scale height of $h_{\rm dust} = 0.10\pm 0.02\,$kpc, so that the
extinction in the foreground of the lens is given by
\begin{equation}
A_{i,L} = {1-e^{-|D_L(\sin b)/h_{\rm dust}|}\over 1-e^{-|D_S (\sin b)/h_{\rm dust}|}} A_{i,S} \ ,
\label{eq-A_L}
\end{equation}
where the index $i$ refers to the passband: $I$, $V$, or $K$. In the Markov Chain calculations themselves,
we fix $D_S = 8.0\,$kpc for our source star at a Galactic longitude of $l = 1.7647$, and we fix the
dust scale height at $h_{\rm dust} = 0.10\,$kpc. But, we remove these restrictions by reweighting the
links in the Markov Chain when we sum them for our final results.

These five constraints have a very small effect on the overall $\chi^2$. The addition of these constraints
increases $\chi^2$ by $\Delta\chi^2 = 1.41$, so it is clear that the light curve is quite consistent with these
constraints.

While these constraints have almost no impact on the best fit model $\chi^2$, they have a dramatic
effect on the allowed range of microlensing parallax parameters, as Figure~\ref{fig-parallax}
indicates. The 2-$\sigma$ range for $\pi_{E,N}$ is reduced from $-0.39 < \pi_{E,N} < 0.43$
to $0.18 < \pi_{E,N} < 0.25$, a reduction of a factor of 12 in uncertainty. This yields a microlensing parallax
amplitude of $\pi_E = 0.265 \pm 0.21$, which will be used in Section~\ref{sec-lens} to determine the 
lens mass.

\section{Lens Properties}
\label{sec-lens}
For most planetary microlensing events, finite source effects provide a measurement of the source radius 
crossing time, $t_*$. This allows the angular Einstein radius, $\theta_E$, to be determined with the 
equation $\theta_E = \theta_* t_E/t_*$, where  $\theta_*$ is the angular source radius, which can be
determined by the source brightness and color \citep{kervella_dwarf,boyajian14}. However, $t_*$,
was not measured for OGLE-2012-BLG-0950, because this event did not reveal any finite source effects.
Fortunately, it is also possible to determine $\theta_E$ from $\bm{\mu}_{\rm rel,G}$, which can be
determined from the measured values of $\bm{\mu}_{\rm rel,H}$ and $\pi_{E,E}$ using 
equations~\ref{eq-mu_helio} and \ref{eq-piE_muG}. The relation between the length of the 
 $\bm{\mu}_{\rm rel,G}$ vector and $\theta_E$ is $\theta_E = t_E \mu_{\rm rel,G}$.
 
The measurement of the either the angular Einstein radius, $\theta_E$, or the microlensing parallax 
amplitude, $\pi_E$, will provide a mass-distance relation, if we assume that the source distance, $D_S$, is
known \citep{bennett_rev,gaudi_araa},
\begin{equation}
M_L = {c^2\over 4G} \theta_E^2 {D_S D_L\over D_S - D_L} 
       =  {c^2\over 4G}{ {\rm AU}\over{\pi_E}^2}{D_S - D_L\over D_S  D_L}  \ .
\label{eq-m_thetaE}
\end{equation}
When both $\theta_E$ and $\pi_E$ are known, the two mass-distance relations in equation~\ref{eq-m_thetaE}
can be multiplied together, yielding
\begin{equation}
M_L = {c^2 \theta_E {\rm AU}\over 4G \pi_E} = {\theta_E \over (8.1439\,{\rm mas})\pi_E} \msun \ ,
\label{eq-m}
\end{equation}
which is a direct mass measurement with no dependence on $D_L$ or $D_S$. 

To solve for the planetary system parameters, we sum over our MCMC results
using the Galactic model employed by \citet{bennett14} as a prior, weighted by the 
microlensing rate and the measured $\mubold_{\rm rel,H}$ value. The lens magnitude measurements
were applied as constraints in the light curve modeling, so we do not apply them again in the
sum over the MCMC results. We do constrain the source distances to follow the microlensing rate
weighted distribution according to our Galactic model, and we evaluate the extinction in the foreground
of the lens using equation~\ref{eq-A_L} with the assumed error bar for $h_{\rm dust}$. The Galactic model is used to consider the uncertainty of the source distance. However, using a fixed source distance of $D_S = 8.2$ kpc does not alter the results. 

\begin{figure}
\epsscale{.9}
\plotone{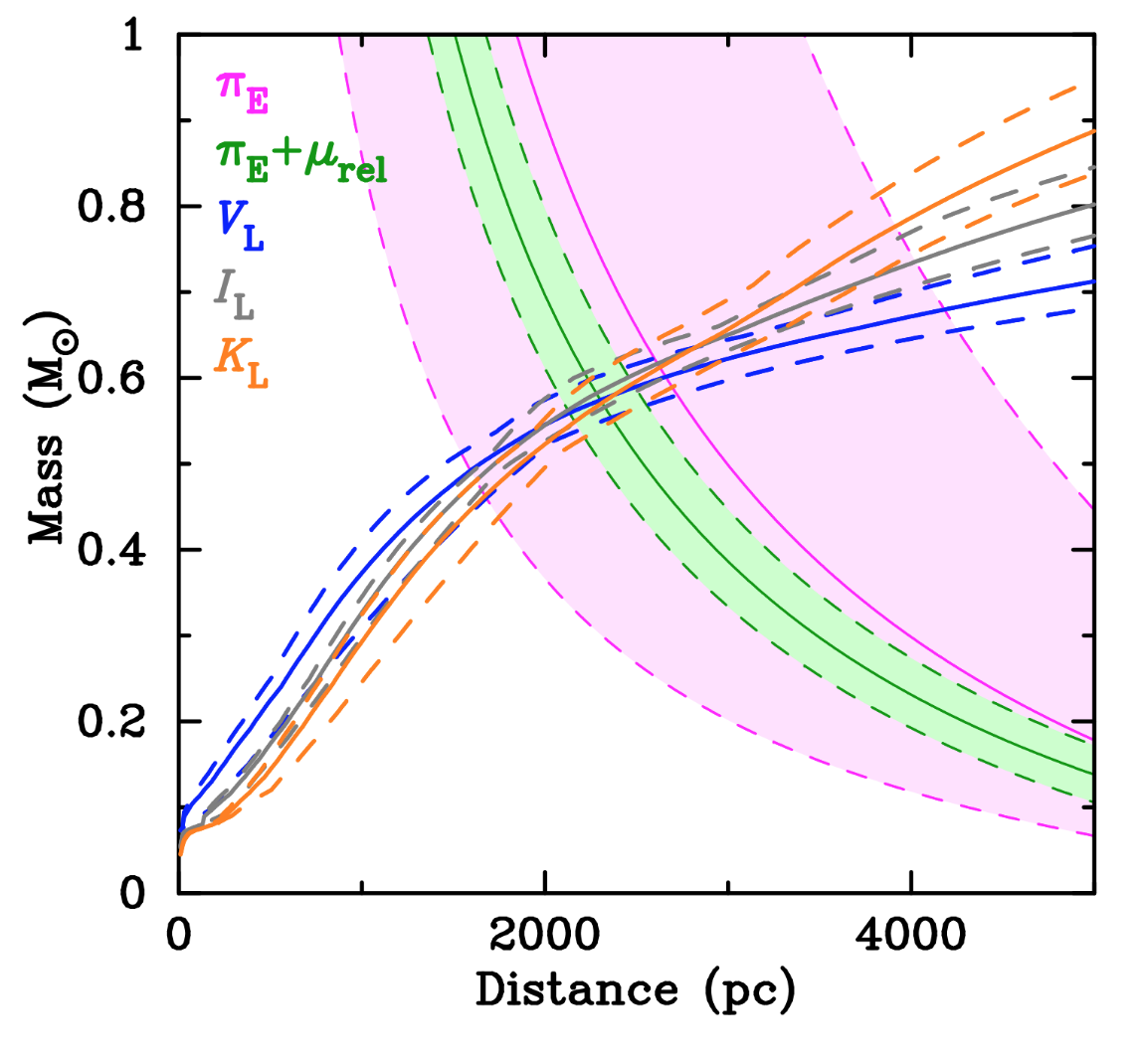}
\caption{ The mass-distance relation obtained from the microlensing parallax parameter determined by 
the light curve models with the $\mubold_{\rm rel,H}$ constraint from the HST and Keck follow-up observations
is plotted in green. The mass-distance relation from the light curve model only, with no additional constraint
on $\pi_{E,N}$ is shaded in magenta. The mass-distance relations obtained from the $K$, $I$, and $V$-band 
mass luminosity relations with the lens flux constraints from Table \ref{tab-dual-fit} are plotted in
red (Keck $K$), black (HST $I$) and blue (HST $V$). The solid lines are best fit values as a function of 
mass and distance. The dashed lines show the 1-$\sigma$ error bars. All three independent flux 
measurements in three different passbands result in the same solution. This confirms our
identification of the lens star.}
\label{fig-mass}
\end{figure}

Figure~\ref{fig-mass} provides a graphical summary of the constraints on the host star in the mass-distance 
plane. The constraint from the one-dimensional microlensing parallax only is the magenta shaded region, 
while the red, black and blue curves give the $K$, $I$ and $V$-band constraints from the Keck and 
HST follow-up observations, with 1-$\sigma$ error bars as dashed lines. Note that a single passband 
flux measurement combined with the one-dimensional parallax constraint does not yield a unique host
star mass. The combination of lens flux constraints in 3 passbands does somewhat better, but it is the
$\bm{\mu}_{\rm rel,H}$ measurement that gives the full ${\bm{\pi}_{\rm E}}$ determination indicated by
the green shaded region. This is the critical feature that provides the precise determination of the 
host star mass, planet mass and lens distance following the method described in section \ref{sec-parallax}. The host mass is measured to be $M_* = 0.58\pm 0.04\msun$, an early M or late K dwarf star, orbited by a twice Neptune-mass planet,
$m_p = 39\pm 8 \mearth$ at a projected separation of $a_{\perp} = 2.54\pm 0.23\,$AU. This also
implies a lens system distance of $D_L = 2.19\pm 0.23$ kpc. The fact that all three excess flux measurements
give the same mass and distance indicates that there is no contamination of measurements from
additional flux from another star \citep{moa310,kosh17_mb16227}.

\begin{figure}
\epsscale{1.0}
\plotone{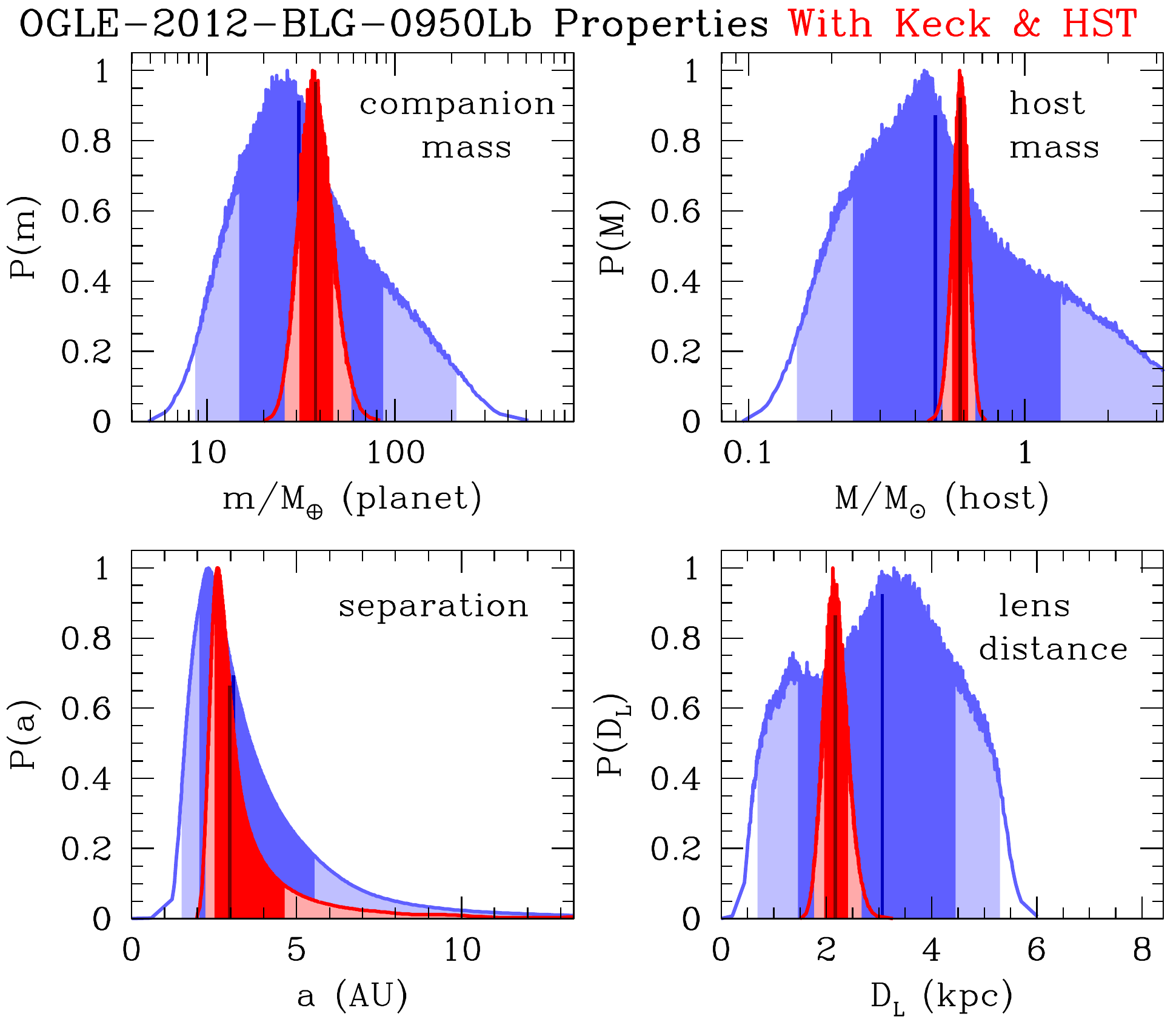}
\caption{The Bayesian posterior probability distributions for the planetary companion mass, host mass, 
their separation and the distance to the lens system are shown with only light curve constraints in blue 
and with the additional constraints from our Keck and HST follow-up observations in red.
The central 68.3$\%$ of the distributions are shaded in darker colors (dark red and dark blue) and the 
remaining central 95.4$\%$ of the distributions are shaded in lighter colors. The vertical black line marks 
the median of the probability distribution of the respective parameters.}
\label{fig-lens}
\end{figure}

\begin{deluxetable}{cccc}
\tablecaption{Measurement of Planetary System Parameters from the Lens Flux Constraints\label{tab-params-histogram}}
\tablewidth{0pt}
\tablehead{\colhead{parameter}&\colhead{units}&\colhead{values \& RMS}&\colhead{2-$\sigma$ range}}
\startdata
Host star mass, $M_*$&${\msun}$&$0.58\pm 0.04$ & 0.51--0.67\\
Planet mass, $m_p$&$M_{\oplus}$& $39\pm 8$& 26--59\\
Host star - Planet 2D separation, $a_{\perp}$&AU&$2.54\pm 0.23$& 2.13--3.03\\
Host star - Planet 3D separation, $a_{3D}$&AU&$3.0^{+1.7}_{-0.5}$& 2.2--10.8\\
Lens distance, $D_L$&kpc &$2.19\pm 0.23$& 1.77--2.68\\
\enddata
\end{deluxetable}

We also use the lens flux and relative lens-source proper motion measurements to constrain the light curve models. The results of our final sum over the Markov Chain light curve models are given in 
Table~\ref{tab-params-histogram} and Figure \ref{fig-lens}. This table gives the mean and RMS
uncertainty plus the central 95.4\% confidence interval range for each parameter except the 
3D separation, $a_{3D}$, where we give the median the central 68.3\% confidence interval.
The lens flux and the parallax measurements 
exclude most of the masses and distances for this planetary system that were compatible with 
Bayesian analysis a MCMC of light curve model without any $\bm{\mu}_{\rm rel,H}$ or lens brightness 
constraints. This constrains the parameters: host star mass, planet mass, their separation and their distance from earth as shown in Table 4. They are consistent with the parameters measured using the empirical mass-luminosity relations described in section \ref{sec-parallax} as described in previous paragraph.
Assuming a random orientation, this implies 3-dimensional separation of $a_{3d} = 3.0 ^{+1.7}_{-0.5}$AU. The uncertainties are calculated as RMS over the MCMC links. 

\section{Discussion and Conclusions}
\label{sec-discussion}
With near simultaneous high angular resolution follow-up observations from Keck and HST, we have 
measured the angular separation of the source and planetary host star to be $34\,$mas in three different
passbands, $K$, $I$, and $V$. This separation measurement allows us to convert the partial measurement
of the microlensing parallax from the light curve into a complete measurement of the two dimensional
${\bm \pi_{\rm E}}$ vector. The combination of this microlensing parallax measurement and the magnitude
of the lens, determined independently in three different passbands. This rules out alternative explanations
of this event involving additional stars such as a companion to the lens or source. The one, highly unlikely
possibility that is not yet excluded is that the detected excess flux could come from a $\sim 0.6\msun$ 
companion to a white dwarf planetary host star that is also $\sim 0.6\msun$. According a statistical study
by \citet{holberg2013}, the probability of a white dwarf hosting a main sequence companion is $\sim$8$\%$, 
but the vast majority of these white dwarf-main sequence star binaries have separations that are much
too large for the main sequence star to be confused with the lens. So, the fraction of white dwarfs with 
binary companions that could be confused with the lens star is about 1\%. However, it is not guaranteed
that a binary companion to a white dwarf lens would have $V$, $I$, and $K$-band magnitudes 
compatible with the lens mass inferred from the $\pi_{\rm E, E}$ measurement and the apparent
$\bm{\mu}_{\rm rel,H}$ measurement, so the probability that a white dwarf host with a binary companion
could produce the light curve and followup data for this event is $\simlt 0.1$\%.
To avoid a light curve signal from
this binary companion, this companion would need to be separated by 
$\simgt 10\theta_E \approx 13\,$mas from the lens at the time of the event. This means
that this unlikely possibility of a white dwarf host star with a main sequence companion could be tested
with an additional epoch of followup observations to confirm that the relative proper motion of the
main sequence star does
extrapolate back to the position of the source at the time of the event, as we have shown for planetary
microlensing event OGLE-2005-BLG-169 \citep{ogle169,batistaogle169}.

The measured planetary mass of $m_p = 39\pm 8\mearth$ is of particular interest because the core accretion
theory predicts that such planets should be rare. The core accretion theory includes a runaway gas accretion 
phase \citep{pollack96,lissauer2009} that is thought to imply that planets in the mass range 20-$80\mearth$
are rare. According to this theory, beyond the snow line, a planetary core rapidly grows by the accumulation 
of planetesimals until it reaches a mass of $\sim 10\mearth$ \citep{pollack96, rafikov11}. Further growth is
dominated by gas accretion that starts slowly, but when the gas mass grows to equal the core mass, but
growth is thought to become a runaway exponential growth process. This process is thought to continue
very rapidly until it is terminated by a lack of gas at a mass similar to that of Jupiter ($318\mearth$) or possibly
Saturn ($95\mearth$). 

The cold exoplanet mass ratio function measured by \citet{suzuki2016} finds no evidence for a dearth of
planets at these intermediate 20-$80\mearth$ masses. The measured mass ratio function increases smoothly
from a mass ratio of $q = 0.03$ down to a mass ratio of $q\approx 10^{-4}$, where it reaches a peak
\citep{udalski18}. There is no evidence of a mass ratio gap at 1-$4\times 10^{-4}$, where we would 
expect to see this expected low occurrence rate of 20-$80\mearth$ mass planets. However, since the
\citet{suzuki2016} considers only mass ratios and not masses, it remains possible that a gap in the
exoplanet mass distribution is smoothed out by the combination of a range of host stars from very low
mass stars up to solar type stars. Perhaps planet formation is different for very low mass stars in a 
way that smooths out the gap in the exoplanet mass distribution around solar type stars. 

Our  follow-up high angular resolution imaging program addresses this issue directly by determining
the masses of the microlens host stars and their planets. The planet OGLE-2012-BLG-0950Lb is
the first planet from the \citet{suzuki2016} sample to have a mass measured to be in the 20-$80\mearth$
range, but our ongoing follow-up observing program will measure masses of more host stars 
and planets from the \citet{suzuki2016} sample to provide a more definitive answer to this question.
Our program has also identified a similar mass planet OGLE-2012-BLG-0026Lb \citep{ogle0026}
with a mass of $46.0\pm 2.5\mearth$ orbiting a solar type star. This planet is one of two planets
detected in this microlensing event that is, unfortunately, not part of the \citet{suzuki2016} sample.

This work is also an important step development in the development of the exoplanet mass measurement
methods \citep{bennett07} for the WFIRST microlensing exoplanet survey \citep{WFIRST_AFTA}. This
requires accurate relative astrometry of the blended source plus lens stars, and in this analysis
we have measured the lens-source relative proper motion to a precision (in the $I$-band) of 2\%, 
despite the fact that the lens and source were separated by $\simlt 0.5\,$FWHM. The 
\citet{moa310} measurement of the separation of a source and blend star $0.17\,$FWHM to a precision 
of 27\% might be considered even more impressive, but this case of OGLE-2012-BLG-0950 involves
a lens star, instead of an unrelated blend, and it has been independently confirmed in three passbands
(instead of only 2).
 
This analysis has also been the first example of a full microlensing parallax measurement being
obtained from light curve measurement of one component of the ${\bm{\pi}_{\rm E}}$ vector and
a follow-up measurement of the Heliocentric lens-source relative proper motion, $\bm{\mu}_{\rm rel,H}$.
This is also an important part of the WFIRST exoplanet mass measurement tool set \citep{bennett07}, 
and in most cases, it will allow for mass measurements that are independent of the flux detected
from the host star. The flux of the lens is also measured in three passbands giving rise to additional redundant consistent 
constraints on the host star and planet masses and distance. The lens and source are not separately resolved in 
any of the images, but still we measure the separation at a high significance. This combination of the 
unresolved lens-source separation measurement and flux measurement in multiple passbands, 
plus 1-d parallax measurement combines all the major WFIRST mass measurement methods, 
as discussed in \citet{bennett07}.
   
We acknowledge the help of Dr. Peter Stetson on operating DAOPHOT and providing us with a 
current version of the code and feedback on our analysis of Keck data. This paper is based, in part,
on observations made with the NASA/ESA Hubble Space Telescope, which is operated by the Association 
of Universities for Research in Astronomy, Inc., under NASA contract NAS 5-26555. 
These observations are associated with program GO-15455. The Keck Telescope observations and analysis
was supported by a 
NASA Keck PI Data Award, administered by the NASA Exoplanet Science Institute. Data presented 
herein were obtained at the W. M. Keck Observatory from telescope time allocated to the National Aeronautics 
and Space Administration through the agency's scientific partnership with the California Institute of Technology 
and the University of California. The Observatory was made possible by the generous financial 
support of the W. M. Keck Foundation.
DPB, AB, and CR  
were also supported by NASA through grant NASA-80NSSC18K0274.


\end{document}